\documentclass[12pt]{article}
\usepackage{epsfig}
\def\iso#1#2{\mbox{${}^{#2}{\rm #1}$}}
\def\h#1{\iso{H}{#1}}
\def\he#1{\iso{He}{#1}}
\def\li#1{\iso{Li}{#1}}
\def\be#1{\iso{Be}{#1}}
\def\b#1{\iso{B}{#1}}
\def\bor#1{\iso{B}{#1}}

\def\pfrac#1#2{\left( \frac{#1}{#2} \right)}

\newcommand\beq{\begin{equation}}
\newcommand\eeq{\end{equation}}
\newcommand\beqar{\begin{eqnarray}}
\newcommand\eeqar{\end{eqnarray}}
\def\ga{\mathrel{\mathpalette\fun >}}
\def\fun#1#2{\lower3.6pt\vbox{\baselineskip0pt\lineskip.9pt
  \ialign{$\mathsurround=0pt#1\hfil##\hfil$\crcr#2\crcr\sim\crcr}}}

\begin{document}
\begin{titlepage}
\pagestyle{empty}
\baselineskip=21pt
\vspace*{-0.6in}
\rightline{astro-ph/0312629}
\rightline{UMN--TH--2226/03}
\rightline{FTPI--MINN--03/38}
\vskip 0.2in
\begin{center}
{\large{\bf Solar Neutrino Constraints on the BBN Production of Li}}
\end{center}
\begin{center}
\vskip 0.2in
{{\bf Richard H. Cyburt}$^{1,2}$, {\bf Brian D. Fields}$^3$
and {\bf Keith A. Olive}$^4$}\\
\vskip 0.1in
{\it
$^1${TRIUMF, Vancouver, BC V6T 2A3 Canada}\\
$^2${Department of Physics, University of Illinois, Urbana, IL 61801, USA}\\
$^3${Center for Theoretical Astrophysics,
Department of Astronomy, \\ University of Illinois, Urbana, IL 61801, USA}\\
$^4${William I. Fine Theoretical Physics Institute, \\
University of Minnesota, Minneapolis, MN 55455, USA}}\\
\vskip 0.2in
{\bf Abstract}
\end{center}
\baselineskip=18pt \noindent
Using the recent WMAP determination of the baryon-to-photon ratio,
$10^{10} \eta = 6.14$ to within a few percent, big bang
nucleosynthesis (BBN) calculations can make relatively accurate
predictions of the abundances of the light element isotopes which can
be tested against observational abundance determinations. At this
value of $\eta$, the $^7$Li abundance is predicted to be significantly
higher than that observed in low metallicity halo dwarf stars.  Among
the possible resolutions to this discrepancy are 1) $^7$Li depletion
in the atmosphere of stars; 2) systematic errors originating from the
choice of stellar parameters - most notably the surface temperature;
and 3) systematic errors in the nuclear cross sections used in the
nucleosynthesis calculations.  Here, we explore the last possibility,
and focus on possible systematic errors in the
$\he3(\alpha,\gamma)\be7$ reaction, which is the only important \li7
production channel in BBN.  The absolute value of the cross section
for this key reaction is known relatively poorly both experimentally
and theoretically.  The agreement between the standard solar model and
solar neutrino data thus provides additional constraints on variations
in the cross section ($S_{34}$).  Using the standard solar model of
Bahcall, and recent solar neutrino data, we can exclude systematic
$S_{34}$ variations of the magnitude needed to resolve the BBN \li7
problem at $\ga 95\%$ CL.  Additional laboratory data on
$\he3(\alpha,\gamma)\be7$ will sharpen our understanding of both BBN
and solar neutrinos, particularly if care is taken in determining the
absolute cross section and its uncertainties. Nevertheless, it already
seems that this ``nuclear fix'' to the
\li7 BBN problem is unlikely; other possible solutions are briefly discussed.

\end{titlepage}
\baselineskip=18pt

\section{Introduction}

The recent all-sky, high-precision
measurement of microwave background anisotropies by WMAP
\cite{wmap} has opened the possibility for new precision analyses of
big bang nucleosynthesis (BBN).  Until now, one could use the
predictions of standard BBN \cite{bbn,sarkar} for the abundances of
the light element isotopes, D, \he3, \he4, and \li7 and compare those
results with the observational determination of those isotopes and
test the concordance of the theory.  If concordance is found, the
theory is also able to predict the value of the baryon-to-photon
ratio, $\eta$.  Indeed, concordance is found, so long as a liberal
estimation of systematic uncertainties are included in the analysis.
The accuracy of the predicted value of $\eta$ from BBN alone based on
likelihood methods \cite{fo,lik,fior,cfo1} is modest: $\eta_{10} =
5.7^{+1.0}_{-0.6}$ when D, \he4, and \li7 are used, and $\eta_{10} =
6.0^{+1.4}_{-0.5}$ when using D alone, where $\eta_{10} = 10^{10}
\eta$.  This pales in comparison with the recent WMAP result of
$\Omega_B h^2 = 0.0224 \pm 0.0009$ which is equivalent to $\eta_{\rm
10,CMB} = 6.14 \pm 0.25$. This result is the WMAP best fit assuming a
varying spectral index and is sensitive mostly to WMAP alone
(primarily the first and second acoustic peaks) but does include CBI
\cite{cbi} and ACBAR \cite{acb} data on smaller angular scales, and
Lyman $\alpha$ forest data (and 2dF redshift survey data \cite{2df})
on large angular scales.

If we use the WMAP data to fix the baryon density, we can make quite
accurate predictions for the light element abundances.  At this
density, we can make a direct comparison \cite{cfo3} between theory
and observation as shown in Table \ref{tab:bbn}.

\begin{table}[htb]
\begin{center}
\caption{Light Element Abundances:  BBN Predictions and Observations
\label{tab:bbn}}
\begin{tabular}{|c|c|c|}\hline\hline
element  & theory    &     Observation\\ \hline
D/H & $ 2.75^{+0.24}_{-0.19} \times 10^{-5}$      &    
$ 2.78 \pm 0.29 \times 10^{-5}$       \\ 
\he4 & $0.2484^{+0.0004}_{-0.0005}$        &     
$0.238 \pm 0.002 \pm 0.005$       \\
\li7 & $3.82^{+0.73}_{-0.60} \times 10^{-10}$ & 
$1.23^{+0.34}_{-0.16} \times 10^{-10}$ \\
\hline\hline
\end{tabular}
\end{center}
\end{table}

As one can see, the agreement between the predicted abundance of D/H
and the observed value (based on the average of the 5 best determined
quasar absorption system abundances \cite{bt,omeara,kirkman,pet}) is
perfect.  The comparison with \he4 is less good, as BBN predicts a
mass fraction which is high compared to most observations
\cite{iz,oss,fo98,peim}.  The value in Table \ref{tab:bbn} is based on a
combined analysis \cite{fo98} which is close agreement with the recent
observations of \cite{peim}.  One should note that 1) the data of
\cite{iz} alone give a higher value for the \he4 abundance $Y_p =
0.242 \pm 0.002 \pm 0.005$, and 2) important systematic effects have
been underestimated \cite{OSk}.  Among the most probable cause for a
serious underestimate of the \he4 abundance is underlying stellar
absorption.  Whether or not this effect can account for the serious
discrepancy now uncovered remains to be seen.

Clearly the key problem concerning the concordance of BBN theory and
the observational determinations of the light element abundances is
\li7.  The primordial abundance of \li7 is determined from the ``Spite
plateau'' \cite{spites} in Li/H observed in low metallicity halo dwarf
stars (extreme Population II).  
The observed value is clearly discrepant with the BBN+WMAP prediction.  
The cause of the
discrepancy may be:

\begin{itemize}
\item Stellar depletion of \li7 -- however, the lack of dispersion in the 
observed data, make it unlikely that dispersion alone can account for
the difference.

\item Stellar parameters -- the determined \li7 is sensitive to the assumed 
surface temperature of the star.  However, to account for a
discrepancy this large temperatures would have to be off by at least
500K. This may not be reasonable.

\item The nuclear rates -- this is the case we wish to explore here.  
\end{itemize}

Of course, it is also possible that the \li7 discrepancy is real, and
points to new physics.  However, it is our view that 
at present, the case for new
physics is not compelling, though it certainly merits serious
investigation.  Furthermore, a firm rejection of the more ``prosaic''
possibilities we have outlined is a prerequisite which must be
satisfied before we are driven to more radical and exciting new
solutions.  It is in this spirit that we investigate possible
systematic errors in the BBN theoretical predictions for \li7.

Uncertainties in the nuclear reaction rates
which determine \li7 are dominated by \he3$(\alpha,\gamma)$\be7.
There has been concerted experimental and theoretical 
effort to understand this reaction, and indeed the
cross section {\em shape} versus energy appears to be
well-understood \cite{Nollett:2001ub}.
However, a challenge to both experimental and theoretical work
has been the determination of the absolute {\em normalization}
of of the cross section.  This uncertainty propagates
into an overall systematic
error in the \he3$(\alpha,\gamma)$\be7 rate.

We thus pose the following question.
Independent of the quoted (or derived) laboratory uncertainties in
\he3$(\alpha,\gamma)$\be7,
what is the maximum allowable amount that this rate can be
adjusted. Of course, we are not completely free to adjust this rate,
since this nuclear reaction occurs in the Sun
and is in part
responsible for the observed flux of solar neutrinos.  Thus our goal
is to use the standard solar model
\cite{bah} as a constraint on the BBN nucleosynthesis rates.  In order
to reduce the predicted \li7 abundance in Table 1, to the observed one
requires a reduction in the \he3$(\alpha,\gamma)$\be7 by a factor of
0.27.  We show that by using the concordance between the standard
solar model and the observed flux of solar neutrino, this is excluded
at the 99.9999 \% CL.  At the 95\% CL, the largest reduction factor
possible is 0.65.  Thus, it is not possible to argue that the
uncertainties in nuclear reactions are solely responsible for the \li7
discrepancy.

In section 2, we detail the problem of BBN produced \li7.  In section
3, we discuss the key nuclear reactions which contribute to the
overproduction of \li7.  We derive our constraints on these reactions
using the observed flux of solar neutrinos in section 4.  A summary
and discussion is given in section 5.

\section{The Overproduction of \li7}  

As noted in Table 1, the BBN \li7 abundance is predicted to be
$3.82^{+0.73}_{-0.60} \times 10^{-10}$ for $\eta_{10} = 6.14 \pm
0.25$.  This result \cite{cfo3} is based on a BBN calculation
\cite{cfo1} using the updated rates compiled by the NACRE
collaboration \cite{nacre}.  Other calculations tend to give even
higher values, e.g., \li7/H = 4.87$^{+0.64}_{-0.60}$~\cite{nb}; \li7/H
= 4.18 $\pm$ 0.46~\cite{cva}.
These results differ due to the different nuclear data sets and
procedures used to fit them and derive thermonuclear rates.
The variations are thus a measure of {\em known} systematics
in the \li7 prediction.
 
 The observed Li/H value in Table 1 reflects the inferred mean
 abundance in the atmospheres for a set of Pop II stars.  The analysis
 is that of \cite{rbofn}, based on the data of \cite{rnb}.  The data
 sample consists of 23 very metal poor halo stars, with metallicities
 ranging from [Fe/H] = -2.1 to -3.3.  The data show a remarkably
 uniform abundance of Li and negligible dispersion about a tiny slope
 which is consistent with the production of some Li in Galactic cosmic
 ray collisions (primarily $\alpha + \alpha$). Note that any Galactic
 component of Li only compounds the BBN discrepancy.
 
The \li7 value in Table 1 assumes that the Li abundance in the stellar
sample reflects the initial abundance at the birth of the star;
however, an important source of systematic uncertainty comes from the
possible depletion of Li over the $\ga 10$ Gyr \cite{sneden} age of
the Pop II stars.  Stellar interiors can burn Li and alter its surface
abundance.  The atmospheric Li abundance will suffer depletion
if the outer layers of the stars have been transported deep enough
into the interior, and/or mixed with material from the hot interior;
this may occur due to convection, rotational mixing, or diffusion.
However, if mixing processes are not efficient, then Li can remain
intact and undepleted in a thin outer layer of the atmosphere, which
contains a few percent of the star's mass but is the portion of the
star's material that is observable.

Standard stellar evolution models
predict Li depletion factors which are very small
($<$0.05~dex) in very metal-poor turnoff stars
\cite{ddk}. However,  there is no reason to 
 believe that such simple models incorporate all effects which lead to
 depletion such as rotationally-induced mixing and/or diffusion.
 Current estimates for possible depletion factors are in the range
 $\sim$~0.2--0.4~dex \cite{dep}.  While the upper end of this range is
 close to the required depletion factor of $\simeq 0.3$ necessary to
 account for the difference in the BBN and observed abundance,
 depletion models typically predict the existence of star-to-star
 differences in observed Li abundances due to the range of stellar
 rotation and other intrinsic stellar properties to which the models
 have some sensitivity. As noted above, this data sample \cite{rnb}
 shows a negligible intrinsic spread in Li leading to the conclusion
 that depletion in these stars is as low as 0.1~dex.

Another important source for potential systematic uncertainty stems
from the fact that the Li abundance is not directly
observed but rather, inferred from an absorption line strength and
a model stellar atmosphere. Its determination
depends on a set of physical parameters and a model-dependent analysis
of a stellar spectrum.  Among these parameters, are the metallicity
characterized by the iron abundance (though this is a small effect),
the surface gravity which for hot stars can lead to an underestimate
of up to 0.09 dex if log g is overestimated by 0.5, though this effect
is negligible in cooler stars.  Typical uncertainties in log g are
$\pm 0.1 - 0.3$.  The most important source for error is the
surface temperature.  Effective-temperature calibrations for stellar
atmospheres can differ by up to 150--200~K, with higher temperatures
resulting in estimated Li abundances which are higher by $\sim
0.08$~dex per 100~K.  Thus accounting for a difference of 0.5 dex
between BBN and the observations, would require a serious offset of
the stellar parameters.

We note however, that a recent study \cite{bon1} with temperatures
based on H$\alpha$ lines (considered to give systematically high
temperatures) yields \li7/H = $(2.19\pm 0.28) \times 10^{-10}$. These
results are based on a globular cluster sample and do show
considerable dispersion.  A related study (also of globular cluster
stars) gives \li7/H = $2.29 \times10^{-10}$ \cite{bon2}.  The
difference between these results and the BBN value is just over 0.2
dex making it plausible that depletion may be responsible for the
difference in these stars which show systematically high temperatures.  
It remains an open question why stars in a globular
cluster--which are usually thought of as sharing a common origin site
and epoch--seem to show a larger Li dispersion (and higher temperatures) 
than field halo stars
whose evolution has not been so tightly related.

Finally, the remaining source of systematic uncertainty pertains not
to the observations, but to the BBN calculation itself.  Here we will
limit ourselves to a discussion of those cross sections which have a
bearing on the production of \be7, which is the dominant source of
mass-7 at the high values of $\eta$ consistent with the WMAP
result.\footnote{At $\eta_{10} = 6.14$, the production ratio is
$\be7/\li7=10.8$.  Of course, the \be7 eventually suffer electron
capture and decay to \li7 before recombination and long before
incorporation into Pop II stars.}  As such the principle cross
sections of interest are: $\he3(\alpha,\gamma)\be7$ and
$\be7(p,\gamma)\b8$.  The reaction $\be7(n,p)\li7$ is not of interest
since it does not largely affect the final abundance of \li7.

\section{Nuclear Rates contributing to BBN \li7 production}

\subsection{Standard BBN}

Since our aim is to fix the \li7 problem by changing nuclear reaction
rates, specifically the $\he3(\alpha,\gamma)\be7$ and
$\be7(p,\gamma)\bor8$ reactions, it is important to understand how
they do or do not impact primordial nucleosynthesis.  We will start
with the all-too-familiar $n(p,\gamma)d$ reaction and how it affects
the light element yields.  This will guide us when looking
specifically at the other reactions.  It is well-known that
nucleosynthesis in the early Universe is delayed due to the deuterium
bottleneck. It is important to understand how the deuterium bottleneck
affects the abundances of the light elements.  The delay being caused
by the large number of photons to baryons, which makes the deuterium
photo-destruction rates much larger than the production rates.  At
lower temperatures, about 70 keV, deuterium production proceeds and
the burning into heavier nuclei occurs until the Coulomb barrier halts
nucleosynthesis.  We burn until we deplete the neutron fuel and the
Coulomb barrier stops charged-induced reactions, happening at a
temperature around 50 keV.

While the bottleneck is in place, neutrons
and protons remain at their weak freeze-out values, except for the
occasional $n$-decay, and deuterium at its equilibrium value.  The
other light element abundances exist in a quasi-static statistical
equilibrium, being determined by various algebraic combinations of the
important thermonuclear reaction rates~\cite{bbnwoc}.  When the
bottleneck ends and the neutron fuel is depleted into \he4, these
abundances tend to freeze-out at particular values depending on the
overall baryon content in the universe and when the bottleneck ended.

The deuterium bottleneck ends when the photo-dissociation rate
$d(\gamma,n)p$ becomes less significant than the $np$-capture rate.
This is done by setting the ratio of equilibrium abundances to unity,
$X_d/X_pX_n\sim1$ for which there is an analytic expression
\cite{bbnwoc} that can be solved iteratively:
\beq
T_d = \frac{B_d}{28.7 - \ln{(\eta_{10}/6.0)} - 1.5\ln{(T_d/{\rm MeV})}}
\eeq
One finds that for $B_d=2.224$ MeV and $\eta_{10} = 6.0$, the
deuterium bottleneck ends at $T_d=0.07$ MeV.  This method can also be
used to determine when certain processes become inefficient, such as
$\be7(\gamma,\alpha)\he3$ and $\b8(\gamma,p)\be7$.  Using equations
that can be similarly solved iteratively, we find the temperature for
which the photo-dissociation of \be7 and \b8 listed earlier is not
significant:
\beqar
T_{34} = \frac{Q_{34}}{28.2 - \ln{(\eta_{10}/6.0)} - 1.5\ln{(T_{34}/{\rm MeV})}} \\
T_{17} = \frac{Q_{17}}{30.1 - \ln{(\eta_{10}/6.0)} - 1.5\ln{(T_{17}/{\rm MeV})}}
\eeqar
Using $Q$-values of 1.587 MeV and 0.137 MeV for the
$\be7(\gamma,\alpha)\he3$ and $\b8(\gamma,p)\be7$ respectively, we
find $T_{34}=0.05$~MeV and $T_{17}=0.004$~MeV.  These temperatures are
below the deuterium bottleneck, suggesting that if these reactions
dominate the formation of these elements, then they should exist at
their equilibrium abundances.  This is true for the
$\he3(\alpha,\gamma)\be7$ reaction.  However, $\be7(p,\gamma)\b8$ must
compete with \b8 decay, with mean lifetime $\sim 1$~s.  An equilibrium
abundance of \b8 during the epoch of big bang nucleosynthesis will be
greatly suppressed and in fact is completely negligible.  In other
words, \b8 and its decay products do not contribute to the primordial
light element abundances.  Knowing that $\he3(\alpha,\gamma)\be7$ is
the dominant contribution to the BBN \li7 abundance prediction, we now
discuss what we need to fix the \li7 problem.

\subsection{Modified BBN:  A Nuclear Solution to the Lithium Problem}

\label{sect:bbn-mod}

The question of interest to us here, is to what extent can these key
rates be altered to enhance the \be7 (\li7) abundance and yet remain
consistent with experimental constraints. To this end, we define a new
$S$-factor\footnote{The $S$-factor is defined by the cross section:
$S(E) = \sigma(E) E \exp(8 \pi^2 \alpha Z_1 Z_2 /v)$. The
last term is the Coulomb penetration factor, in which $Z_i$ are the
charges of the incoming nuclei and $v$ their relative velocity.}
$S_{17}^{NEW}$ which we assume for simplicity to be proportional to
the old one, $S_{17}^{OLD}$.  Note that for $S_{17}$, a
proportionality factor between 0 and 2 does not change the BBN
predictions significantly. In contrast, the dependence of the mass 7
abundance on $S_{34}$ is nearly 1:1, as apparent in Table
\ref{tab:bbn_alt}, and in good agreement with the results of
\cite{fior}, who find that $\li7_{\rm BBN} \propto S_{\rm 34}^{0.95}$.

\begin{table}[htb]
\begin{center}
\caption{\li7 Sensitivity to $S_{34}$
\label{tab:bbn_alt}
}
\begin{tabular}{||c|c||}\hline\hline
$S_{34}^{NEW}/S_{34}^{OLD}$    &     Y$_7^{NEW}$/Y$_7^{OLD}$ \\ \hline
1.00    &    1.00     \\ 
0.75    &    0.76     \\
0.50    &    0.52     \\ 
0.25    &    0.28     \\ 
\hline\hline
\end{tabular}
\end{center}
\end{table} 

As discussed above, there are two sets of \li7 observations we can try to
match by renormalizing the $\he3(\alpha,\gamma)\be7$ reaction.
Using the \li7 measurements of a metal poor globular cluster \cite{bon2}
would require a change of $S_{34}^{NEW} =
0.53S_{34}^{OLD}$.  Using the \li7 measurements of metal poor stars in
the Galactic halo \cite{rbofn} would require a change of
$S_{34}^{NEW}=0.27S_{34}^{OLD}$.  

The determination of the BBN light element yields is from 
\cite{cfo1}, where new normalizations
and errors to the NACRE~\cite{nacre} rates important for primordial
nucleosynthesis  have been assigned.  
The value, $S_{17}^{OLD}(0)=0.021\pm 0.002$ keV b is
taken straight from the NACRE collaboration.  For
$\he3(\alpha,\gamma)\be7$, the BBN calculation uses the renormalized
NACRE rate $S_{34}^{OLD}(0)=0.504\pm 0.0534$ keV b \cite{cfo1}.  
As one can see, shifts in the
$\he3(\alpha,\gamma)\be7$ cross section as large as that necessary to
produce $S_{34}^{NEW}$ are strongly excluded given the cited
uncertainties for this reaction.  Although adjustments in the nuclear
cross-sections of this size are unlikely given the stated experimental
errors, one could worry that additional systematic effects are
present, particularly given the difficulties in establishing the
absolute normalization for this reaction.  As stated in the
Introduction, these rates in particular can be bounded by another
means. In the next section, we will determine the maximum possible
downward adjustment to $S_{34}$ which is consistent with solar
neutrino fluxes.

The effect of changing the yields of certain BBN reactions was
recently considered by Coc et al. \cite{cva}.  In particular, they
concentrated on the set of cross sections which affect \li7 and are
poorly determined both experimentally and theoretically.  In many
cases however, the required change in cross section far exceeded any
reasonable uncertainty.  Nevertheless, it may be possible that certain
cross sections have been poorly determined.  In \cite{cva}, it was
found for example, that an increase of the $\li7(d,n)2\he4$ reaction
by a factor of 100 would reduce the \li7 abundance by a factor of about
3.  Another reaction which is poorly determined is $\be7(d,p)2\he4$.
An increase in this rate by a factor of $\sim 100$ could also
alleviate the \li7 discrepancy

\section{The Sun as a Nuclear Laboratory}

The $\he3(\alpha,\gamma)\be7$ reaction plays a crucial role
not only in BBN \li7 synthesis, but also in solar neutrino
production.  In particular, this reaction is responsible for
the creation of \be7, which will then either (1) 
produce a monoenergetic neutrino
via electron capture $\be7 (e^-,\nu_e) \li7 $, or (2) 
produce \bor8 via radiative capture of a proton, $\be7(p,\gamma)\bor8$.
The branching between these paths determines the solar
\bor8 abundance and thus directly sets the flux of
\bor8 neutrinos.  SNO (as well as Super-K) are
sensitive exclusively to these neutrinos.  Furthermore, SNO measures
directly the total \b8 neutrino flux, with no assumptions about
mixing~\cite{sno}.  They find:
\beq
\label{eq:sno}
\phi_8 = \left[5.21\pm0.27(stat)\pm0.38(syst)\right]\times10^6 \ \ {\rm cm^{-2}
\ s^{-1}},
\eeq
where this is determined with no assumed shape of the
\b8 energy spectrum. 
This flux thus offers a constraint on the $\he3(\alpha,\gamma)\be7$
reaction, as follows.

The Standard Solar Model of Bahcall \cite{bahcall}
can be used to predict the solar
neutrino fluxes that can be observed by experiments.  These fluxes
depend upon various solar parameters, such as the luminosity, the
chemical abundances, and nuclear fusion cross sections.  In fact, the
neutrino flux uncertainties are dominated by the cross section errors.
Provided by \cite{bahcall}, simple scalings between neutrino fluxes
and these cross sections robustly describe the SSM predictions.  The
\b8 neutrino flux scaling is:
\beq
\label{eqn:8scalings}
\phi_8 \propto S_{11}^{-2.6}S_{33}^{-0.4}S_{34}^{0.81}S_{17}^{1.0}S_{e7}^{-1.0}.
\eeq
Here, the $S$'s are the astrophysical $S$-factors, except for
$S_{e7}$.  The $S_{e7}$ reaction is the electron capture rate on
\be7. One usually takes some nuclear rate compilation and uses the
$S$-factors to evaluate the neutrino flux given these scalings.  Two
such nuclear compilations are from Adelberger {\em et
al}~\cite{adelberger} and the NACRE collaboration~\cite{nacre}.  Their
determinations relevant for this work are shown in
Table~\ref{tab:rate}.

\begin{table}[ht]
\begin{center}\caption{Shown are the results from the nuclear fusion rate
compilations.}
\label{tab:rate}
\begin{tabular}{||l|l|l||}
\hline\hline
Reaction & Adelberger~\cite{adelberger} (keV b) & NACRE~\cite{nacre}
(keV b) \\
\hline\hline
$p(p,e^{+}\nu_e)\h2$ & $S_{11} =
4.0\times10^{-22}(1.0\pm0.007^{+0.020}_{-0.011})$ & $S_{11} =
3.94\times10^{-22}(1.0\pm0.05)$\\
\hline
$\he3(\he3,2p)\he4$ & $S_{33} = 5.4\times10^3(1.0\pm0.074)$ & $S_{33} =
5.18\times10^3(1.0\pm0.06)$ \\
\hline
$\he3(\alpha,\gamma)\be7$ & $S_{34} = 0.53(1.0\pm 0.09434)$ & $S_{34} =
0.54(1.0\pm 0.167)$ \\
\hline
$\be7(p,\gamma)\b8$ & $S_{17} = 0.019(1.0^{+0.21}_{-0.11})$ & $S_{17} =
0.021(1.0\pm0.11)$ \\
\hline
$\be7(e^{-},\nu_e)\li7$ & $S_{e7} = 5.6\times10^{-9}(1.0\pm0.02)$ s$^{-1}$ & -NA- \\
\hline\hline
\end{tabular}
\end{center}
\end{table}

Our strategy is thus to use the SNO measurements of the \bor8 neutrino
flux and the SSM to constrain $S_{34}$.  This is accomplished via the
scalings in eq.\ (\ref{eqn:8scalings}).  A complication is that these
scalings also depend on other reactions, none of which are significant
for BBN, and all of which are better measured than
$\he3(\alpha,\gamma)\be7$.  This approach amounts to the extreme case
in which we ignore {\em all} of the hard-won laboratory and
theoretical information on $S_{34}$, using only solar neutrino data as
well as constraints on other reactions, $S_{17}$.  This can be viewed
as providing independent information about $S_{34}$, or as a test of
the systematics in the normalization, which is a salient feature for the
BBN \li7 problem.  Our results will thus use the Sun to provide new
and independent limits on the systematics of $S_{34}$.  We will derive
these using both approximate analytical methods and more accurate
numerical methods.

\subsection{Analytic Formalism and Results}
We can estimate the impact these rate compilations have on the
neutrino flux, by doing linear error propagation as follows:
\beq
\left( \frac{\sigma_8}{\phi_8}\right)^2 \approx
\left(\frac{2.6\sigma_{11}}{S_{11}}\right) ^2 + \left(
\frac{0.4\sigma_{33}}{S_{33}} \right) ^2 + \left( \frac{0.81\sigma_{34}}{S_{34}}
\right) ^2 + \left( \frac{\sigma_{17}}{S_{17}} \right) ^2+ \left(
\frac{\sigma_{e7}}{S_{e7}} \right)^2.
\eeq
We find that the Adelberger and NACRE compilations predict
$\sigma_8/\phi_8 = \pm0.19$ and $\sigma_8/\phi_8=\pm0.22$ respectively
using this linear approximation.  With these results, we find that the
error in the predicted flux is determined primarily by the $S_{17}$,
$S_{34}$ and $S_{11}$ reactions.  With our ultimate aim of
constraining $S_{34}$, we will have to treat at least the $S_{17}$ and
$S_{11}$ uncertainties directly, in addition to the error in the solar
neutrino flux measurement.

We can now use the scalings in eqn.~\ref{eqn:8scalings} to estimate
the likely value of $S_{34}$ and its uncertainty, based on the SNO
observations (\ref{eq:sno}),
\beq
\label{eq:34scalings}
\frac{S_{34}}{S_{34,0}} =
   \pfrac{S_{11}}{S_{11,0}}^{3.21}
   \pfrac{S_{33}}{S_{33,0}}^{0.49} 
   \pfrac{S_{17}}{S_{17,0}}^{-1.23} 
   \pfrac{S_{e7}}{S_{e7,0}}^{1.23} 
   \pfrac{\phi_8}{\phi_{8,0}}^{1.23}
\eeq
where we use the Bahcall et al.\ results for
the Adelberger and NACRE reaction complications (the $S_{i,0}$)
to determine the flux normalization
\beqar
\phi_{8,0}^{\rm ADL} & = &  5.05 \times 10^{-6} {\rm cm^{-2} \ s^{-1}} \\
\phi_{8,0}^{\rm NAC} & = &  5.44 \times 10^{-6} {\rm cm^{-2} \ s^{-1}} 
\eeqar
These normalizations are both in excellent agreement with the observed
flux (eq.\ \ref{eqn:8scalings}).  In the extreme case in which {\em
all} of the small mismatch between predicted and observed fluxes is
attributed to $S_{34}$, we expect a shift of
$S_{34}/S_{34}^{ADL}=1.04$ and $S_{34}/S_{34}^{NAC}=0.95$ using the
purely Adelberger and NACRE rate compilations, respectively; the
smallness of these shifts just restates the success of the SSM
in light of the SNO observations.

If we adopt the scaling laws and propagate the errors according to the
usual rules, we have
\beq
\pfrac{\sigma_{34}}{S_{23}}^2
 \approx \pfrac{3.21\sigma_{11}}{S_{11}}^2
   + \pfrac{0.49\sigma_{33}}{S_{33}}^2
   + \pfrac{1.23\sigma_{17}}{S_{17}}^2
   + \pfrac{1.23\sigma_{e7}}{S_{e7}}^2
   + \pfrac{1.23\sigma_8}{\phi_8}^2.
\eeq
This gives a dispersion of $\sigma_{34}/S_{34} = 0.24$ for both
compilations.  This is much larger than the small shifts in the mean
found in the above paragraph.  Moreover, we see that to solve the BBN
\li7 problem with reaction rate uncertainties alone requires a $\sim
2\sigma$ change in $S_{34}$.  Thus we find that this solution to the
\li7 problem is excluded at $\sim 95\%$ CL.  We now turn to numerical
results which will confirm and better quantify this limit.

\subsection{Numerical Formalism and Results}
Our analytic discussion uses standard error propagation which is good
only to first order, and assumes gaussian errors as well as linearity.
To explore this scenario more rigorously, we perform this calculation
numerically, taking into account the non-gaussian nuclear errors and
non-linear scalings.  We set out to perform a Monte Carlo integration
of an integral of the form:
\beq
\int {\mathcal L}_{SSM}(\vec{S},\phi_8) {\mathcal L}_{NUC}(\vec{S}) {\mathcal
L}_{SNO}(\phi_8) d\vec{S} d\phi_8,
\eeq
where $\vec{S}$ is a set of reaction rates (such as the rates already
listed) and $\phi_8$ is the \b8 solar neutrino flux.  ${\mathcal
L}_{SSM}$, ${\mathcal L}_{NUC}$, and ${\mathcal L}_{SNO}$ are the
likelihood distributions of the Standard Solar Model given a reaction
network and a solar neutrino flux, the reaction network given various
rate compilations, and the total \b8 solar neutrino flux given by the
SNO collaboration.

In order to test the reliability and accuracy of this method, we first
predict the total \b8 neutrino flux given a complete reaction network, using
both the Adelberger and NACRE compilations and then compare to the
predictions shown in the works of Bahcall {\em et al} \cite{bah}.  The
integral we are performing is:
\beq
{\mathcal L}(\phi_8) = \int {\mathcal L}_{SSM}(\vec{S},\phi_8) {\mathcal
L}_{NUC}(\vec{S}) d\vec{S}.
\eeq
A Monte Carlo integration uses one of the likelihood functions to draw
random numbers and average the remaining function over those generated
random numbers.  For our case, we will generate random numbers for the
independent reaction rates given by either the Adelberger or NACRE
compilations.  We combine statistical and systematic uncertainties by
adding them in quadrature, since only the total uncertainty is needed
for this analysis.  We generate gaussian or piecewise gaussian
distributions for the reaction rates, depending on whether the quoted
errors are symmetric or asymmetric about the most likely value. For
each random draw of the reaction rates, we can calculate a solar
neutrino flux, given the scalings shown in eqn.~\ref{eqn:8scalings}.
Once a large sample of $\phi_8$ is created, we can calculate its
likelihood distribution.  To summarize:
\begin{enumerate}
\item 
${\mathcal L}_{NUC}(\vec{S})$ generates reaction rates randomly.

\item
${\mathcal L}_{SSM}(\vec{S},\phi_8)$ enforces the scalings in
eqn.~\ref{eqn:8scalings}.

\item
The resulting sample of $\phi_8$ is used to find ${\mathcal L}(\phi_8)$.
\end{enumerate}

The normalization or best value and the errors are
calculated separately.  The flux values for Adelberger and NACRE, as
given in Tables 7 and 9 in \cite{bah} are the standard solar model
predictions for the neutrino fluxes, adopting each compilations best
fit values, without marginalizing over the reaction network.  The
errors are then propagated separately, as described in \cite{bahulr}
using the scalings already mentioned.  The scalings are valid in
determining the uncertainties to 10\%.  Thus, we will adopt the
scalings shown in eqn.~\ref{eqn:8scalings}, normalized such that when
a given compilation is used, we reproduce the values listed in \cite{bah}
\beqar
\phi_8^{ADL} = 5.05\times10^{6} \left(
\frac{S_{11}}{S_{11,0}^{ADL}}\right)^{-2.6} \left(
\frac{S_{33}}{S_{33,0}^{ADL}}\right)^{-0.40} \left(
\frac{S_{34}}{S_{34,0}^{ADL}}\right)^{0.81} \left(
\frac{S_{17}}{S_{17,0}^{ADL}}\right)^{1.0}  \left(
\frac{S_{e7}}{S_{e7,0}^{ADL}}\right)^{-1.0} \\
\phi_8^{NAC} = 5.44\times10^{6} \left(
\frac{S_{11}}{S_{11,0}^{NAC}}\right)^{-2.6} \left(
\frac{S_{33}}{S_{33,0}^{NAC}}\right)^{-0.40} \left(
\frac{S_{34}}{S_{34,0}^{NAC}}\right)^{0.81} \left(
\frac{S_{17}}{S_{17,0}^{NAC}}\right)^{1.0}  \left(
\frac{S_{e7}}{S_{e7,0}^{ADL}}\right)^{-1.0}
\eeqar
By using the Adelberger scaling relation to predict the NACRE scaling
relation and vice versa, we can verify the accuracy of these fits.  We
find deviations from the relations listed above at the 8 or 9\% level,
thus we adopt an overall 10\% systematic uncertainty in the predicted
flux.  Also, since the resulting distributions are non-gaussian, we
expect our marginalized best fit neutrino fluxes to be different from
the neutrino flux determined by adopting only the best values of the
reaction rates.

\begin{figure}
\begin{center}
\epsfig{file=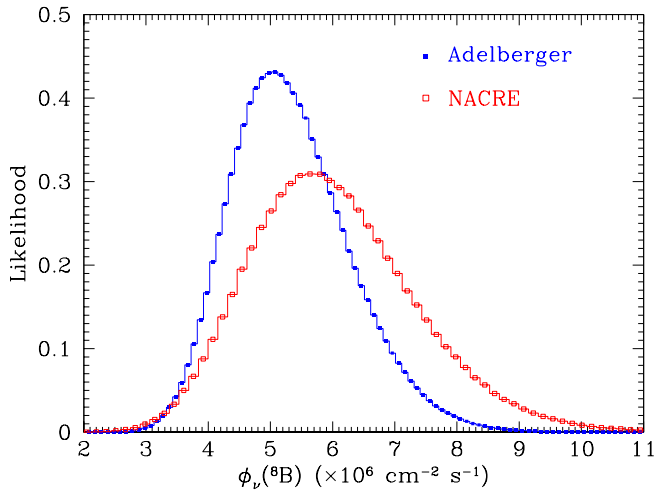, height=4.0in,angle=270}
\end{center}
\caption{Shown are the Standard Solar model predictions of the 
total \b8 neutrino flux.  The binned likelihood based on the
Adelberger (NACRE) rate compilation is plotted with solid (open)
squares.  The 10\% systematic error has not been included here.}
\label{fig:Lphi}
\end{figure}

We find remarkable agreement between our confidence intervals and
those placed by Bahcall {\em et al} \cite{bah}.  Our results are
summarized below, as well as in figure~\ref{fig:Lphi};
\beqar
\phi_8^{ADL} &=& 5.09\left[ 1.0^{+0.20\, (0.44)}_{-0.16\, (0.29)}(stat) \pm 0.10
(syst)\right]\times10^6 {\rm \ cm^{-2} \ s^{-1}} \\
\phi_8^{NAC} &=& 5.19\left[ 1.0^{+0.25\, (0.53)}_{-0.21\, (0.38)}(stat) \pm 0.10
(syst)\right]\times10^6 {\rm \ cm^{-2} \ s^{-1}},
\eeqar
where the flux numbers listed are the most likely values for the
Adelberger-based and NACRE-based compilations and their respective
68\% (95\%) confidence limits, as determined from the marginalized
likelihood distributions.  Notice that our most likely values are
different than the fluxes determined by adopting the best values for
the reaction rates.  This shift in best values is due to the
marginalization over the non-linear scalings and asymmetric nuclear
errors.  Had the scalings been linear and additive, and all nuclear
errors symmetric, no shift would have been seen.  As one expected,
NACRE has slightly inflated errors as compared to Adelberger.  This is
simply due to NACRE's overall larger rate uncertainties, as shown in
our analytic work.

As discussed earlier, the solar neutrino flux depends primarily on the
$S_{17}$ and $S_{34}$ reactions.  The $S_{33}$ and $S_{e7}$ reactions
have little impact on the results due to their small errors and the
weak dependence of the flux on them.  The $S_{11}$ rate has negligible
effect in the Adelberger compilation, but has significant impact in
the NACRE compilation's results.  NACRE's uncertainty for this rate is
a factor of 2 larger than the Adelberger's compilation.  Below we will
use the differing results of these two compilations as a probe of the
$S_{11}$ error assignment.

\begin{figure}
\begin{center}
\epsfig{file=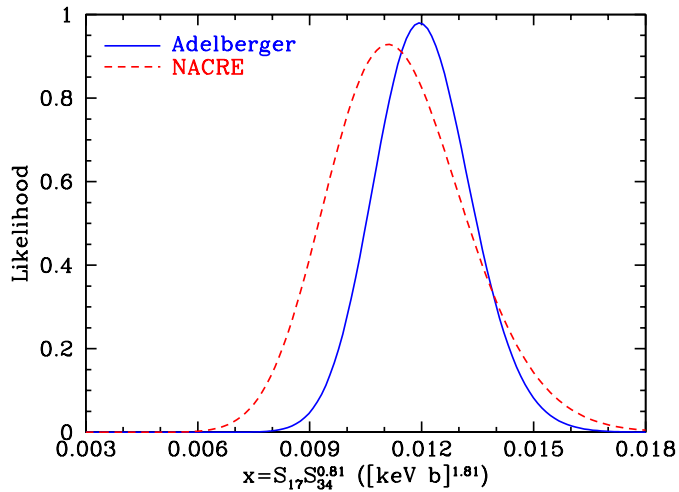, height=4.0in,angle=270}
\end{center}
\caption{Shown are the likelihood distributions of the parameter 
$x=S_{17}S_{34}^{0.81}$, given the subset of reactions from the
Adelberger (solid) and NACRE (dashed) compilations respectively and
the SNO collaborations measurement of the solar \b8 neutrino flux.
The 10\% systematic uncertainty in the flux scalings has been
included as gaussian.}
\label{fig:Lx}
\end{figure}

Given the scalings in eqn.~\ref{eqn:8scalings}, we can use the SSM and
the SNO measurement of the total \b8 neutrino flux to constrain these
rates in the following combination: $x=S_{17}S_{34}^{0.81}$.  As
before, we will generate random numbers for the independent reaction
rates given by either the Adelberger or NACRE compilations.  However,
we now fix $S_{17}$ and $S_{34}$ with various values of $x$.  For each
random draw of the reaction rates, we can calculate a solar neutrino
flux, given the scalings shown in eqn.~\ref{eqn:8scalings}.  With this
flux, we then average ${\mathcal L}_{SNO}(\phi_8)$ over the sample to
find the likelihood of a given $x$.  To summarize:
\begin{enumerate}
\item 
${\mathcal L}_{NUC}(\vec{S})$ generates reaction rates randomly.

\item
${\mathcal L}_{SSM}(\vec{S},\phi_8)$ enforces the scalings in
eqn.~\ref{eqn:8scalings}.

\item
Calculate ${\mathcal L}(x) \equiv \langle{\mathcal L}_{SNO}(\phi_8)\rangle$.
\end{enumerate}
Using the $S_{11}$ and $S_{33}$ from the Adelberger and NACRE
compilations respectively and the $S_{e7}$ from Adelberger
compilation, and the SNO collaborations constraint on the total \b8
neutrino flux, we place the following constraints on $x$.
\beqar
x^{ADL} &=& 0.0119 \left[ 1.0^{+0.16\, (0.35)}_{-0.14\, (0.25)}\right]\ \ [{\rm
keV \ b}]^{1.81}  \\
x^{NAC} &=& 0.0121\left[ 1.0^{+0.21\, (0.46)}_{-0.17\, (0.32)}\right]\ \  [{\rm
keV \ b}]^{1.81}, 
\eeqar
where the most likely values, the 68\% (95\%) confidence
intervals. The 10\% systematic error has been included in the
calculation and assumed to be gaussian.  These resulting likelihoods
for $x$ are shown in figure~\ref{fig:Lx}.  The Adelberger and NACRE
-based results agree quite well with each other.  With differences
mainly attributable to the larger error in $S_{11}$ adopted by NACRE.

\begin{table}[ht]
\begin{center}\caption{Shown are the constraints placed on 
$S_{34}$ using reaction rates from various sources.  Column~1 lists
the adopted $S_{17}$ constraint used, while Columns~2 and 3 show the
compilation used for the $S_{11}$ and $S_{33}$ reaction rates.  The
$S_{34}$ numbers cited are the most likely values and their 68\%
(95\%) confidence intervals.}
\label{tab:S34}
\begin{tabular}{||l|l|l||}
\hline\hline
Adopted $S_{17}$ (eV b) & Adelberger-based~\cite{adelberger} &
NACRE-based~\cite{nacre}\\
\hline\hline
Adelberger~\cite{adelberger} &  & \\
$S_{17}=19.0^{+4.0}_{-2.0}$ & $S_{34}=0.51^{+0.15\, (0.34)}_{-0.12\, (0.21)}$ &
N.A. \\
\hline
NACRE~\cite{nacre} & & \\
$S_{17}=21.0\pm2.31$ & N.A. & $S_{34}=0.51^{+0.17\, (0.38)}_{-0.12\, (0.22)}$ \\
\hline
Junghans~\cite{junghans} & & \\
$S_{17}=21.4\pm0.5(expt)\pm0.6(theor)$ & $S_{34}=0.48^{+0.10\,
(0.23)}_{-0.08\,(0.15)}$ & $S_{34}=0.49^{+0.14\, (0.30)}_{-0.11\, (0.19)}$ \\
\hline
Davids~\cite{davids} & & \\
$S_{17}=18.6\pm0.4(expt)\pm1.1(extrp)$ & $S_{34}=0.57^{+0.13\, (0.30)}_{-0.11\,
(0.19)}$ & $S_{34}=0.59^{+0.17\, (0.39)}_{-0.13\, (0.24)}$ \\
\hline\hline
\end{tabular}
\end{center}
\end{table}
\begin{figure}
\begin{center}
\epsfig{file=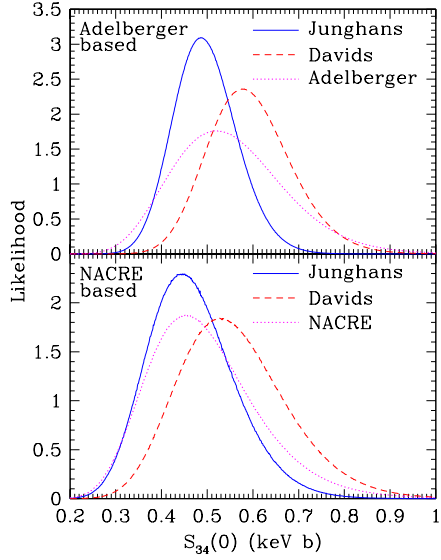, height=3.5in,angle=0}
\end{center}
\caption{Shown are the likelihood distributions of $S_{34}$, given 
$S_{17}$ measurements listed in Table~\ref{tab:S34}.  The upper
(lower) panel shows the results using the Adelberger (NACRE)
compilation for the $S_{11}$, $S_{33}$ and $S_{e7}$ reactions.  We
have used values for $S_{17}$ from Junghans~\cite{junghans} (solid),
Davids~\cite{davids} (dashed) and Adelberger~\cite{adelberger} and
NACRE~\cite{nacre} (dotted).  Again, the 10\% systematic uncertainty
in the scalings has been included and assumed gaussian.}
\label{fig:LS34}
\end{figure}

Since we are constraining $x$ only, we cannot determine the $S_{17}$
and $S_{34}$ reactions uniquely.  We require additional information.
If a total $p-p$ or \be7 neutrino flux measurement existed, we could
in principle determine both cross sections.  Since we are using the
Sun to constrain systematic errors in the normalization of $S_{34}$,
in an attempt to fix the BBN \li7 problem, we will adopt various
experimentally-determined values of $S_{17}$ to place constraints on
$S_{34}$.  Once a value of $S_{17}$ is adopted, we convolve the $x$
likelihood distribution with the experimental $S_{17}$ distribution to
get our $S_{34}$ likelihood.

Besides using the Adelberger and NACRE rate compilations for $S_{17}$,
we also use two more recent determinations.  We use the recommended
values from Junghans {\em et al}.~\cite{junghans}, and Davids and
Typel~\cite{davids}.  The Junghans quoted value,
$S_{17}=21.4\pm0.5(expt)\pm0.6(theor)$ eV b, is based on several
direct capture data sets.  The Davids and Typel value, $S_{17} =
18.6\pm0.4(expt)\pm1.1(extrp)$ eV b, is based on both direct capture
and Coulomb dissociation measurements, excluding the Junghans data set
because it is systematically higher than the other data sets.  Had the
Junghans data been used, the value of $S_{17}$ would lie between the
two cited values.  We will adopt the cited numbers, keeping in mind
that the difference in their values are a measure of this systematic
difference.

Our constraints in Table~\ref{tab:S34} are based on the likelihood
functions in figure~\ref{fig:LS34}.  We find that,
\beq
\label{eq:S34-ssm}
S_{34} > 0.35 {\rm \ keV\  barn}
\eeq
at 95\% CL for the case of the NACRE $S_{17}$ value. Other choices give
slightly higher limits, e.g., Adelberger with the Davids $S_{17}$
gives $S_{34} > 0.42 \ {\rm keV \ barn}$.

As shown in Table \ref{tab:bbn_alt}, these limits on $S_{34}$ place
essentially identical limits to \li7 production in BBN.  Thus, eq.\
(\ref{eq:S34-ssm}), along with the fiducial BBN results in Table
\ref{tab:bbn}, demands that
\beq
\pfrac{\li7}{\rm H}_{\rm BBN} > 2.72^{+0.36}_{-0.34} \times 10^{-10},
\eeq
where we have fixed the reaction normalization such that $S_{34}=0.35$
keV barn, but propagated the other nuclear uncertainties in the BBN
code~\cite{cfo1} and convolved the predictions with the WMAP
determination of the baryon density~\cite{wmap}.  We see that this
alleviates the BBN \li7 problem somewhat, but still requires a
combination of effects to fix the problem--i.e., that \li7 observations
be systematically low, in addition to adopting the limits to nuclear
systematics we have derived.

Put another way, we can ask how far a ``nuclear-only'' BBN solution
stretches our constraints on $\he3(\alpha,\gamma)\be7$ systematics.
We saw in \S\ref{sect:bbn-mod} that for halo star observations to
reflect the primordial \li7 abundance requires that $S_{34}$ be
systematically lowered, to 53\% and perhaps 27\% of its fiducial
value.  A reduction of $S_{34}^{\rm new} < 0.267$ keV barn is excluded
at the 99.5\% CL for the NACRE case (and above for others in Table
\ref{tab:S34}). A reduction of $S_{34}^{\rm new} < 0.136$ keV barn is
excluded at more than 99.9999\% CL.  This restates our finding that
the solar constraints on $S_{34}$ remove this reaction as the main
suspect in the \li7 problem.

\section{Discussion and Conclusions}

The hot big bang cosmology has seen a great triumph in the agreement
between the baryon density found by WMAP and the BBN value implied by
the D/H ratio measured at high redshifts.  However, this triumph is
somewhat muted by the much poorer agreement between the primordial
\li7 value as predicted from BBN and the WMAP baryon density, and the
observed values seen in halo stars.  The predictions are at least a
factor of 2 above the observations. This discrepancy impels a search
for any possible systematic errors, which could either explain the
mismatch, or if no systematics can be found, would reveal the true
seriousness of the problem and a need for a more fundamental solution.

In this paper we have considered the effect of systematic errors in
the nuclear reactions.  In particular, we have focused on the
$\he3(\alpha,\gamma)\be7$ reaction, which is the sole important
production channel of
\li7 at the WMAP baryon density.  As such, systematic
errors in this reaction have an immediate impact on the BBN \li7
abundance. And indeed, while there has been extensive and careful work
for this reaction, both fronts meet with technical difficulties which
leave open the possibility for systematic errors 
in the absolute normalization of this rate.

Thus we have identified a new constraint on this reaction, coming from
its influence on \be7 and \bor8 production in the Sun, and the
associated \bor8 solar neutrinos.  The excellent agreement between the
standard solar model and the total measured \bor8 neutrino flux places
demands that the underlying nuclear reactions cannot have large
systematics.  In particular, using the solar neutrino theory and
observations, as well as some information on other reactions, notably
$\be7(p,\gamma)\bor8$, we find that $S_{34}$ cannot be smaller than
65\% of its fiducial value (e.g., NACRE or Adelberger).  This limit is
strong enough to exclude the $\he3(\alpha,\gamma)\be7$ reaction as the
dominant solution to the BBN \li7 problem.

Other nuclear solutions to the \li7 problem are logically possible but
in fact unlikely.  While many reactions are important for \li7
production, the requirements that we not spoil agreement with D, and
not (further) underproduce \he4, leads us to focus on reactions which
only affect \li7.  Since we have shown that the production channel
cannot be lowered sufficiently, we might hope to increase \be7
destruction.  This is done via the $\be7(n,p)\li7$ reaction, followed
by $\li7(p,\gamma)2\he4$.  The Sun does not constrain $\be7(n,p)\li7$
because the solar interior has a negligible neutron density.  However,
this reaction is nevertheless very well-studied because its inverse is
a common laboratory neutron source.  Since deuterium observations and CMB
determinations suggest a baryon density on the high side, the
destruction of \li7 through the reaction $\li7(p,\gamma)2\he4$ has
negligible impact.  Its mirror reaction, $\be7(n,\gamma)2\he4$,
important on the higher baryon density side, is negligible compared to
$\be7(n,p)\li7$.  Furthermore, \li7 has a somewhat weaker dependence
on the destruction cross section ($\li7_{\rm BBN} \propto S_{\rm
34}^{0.95} S_{n7}^{-0.74}$ \cite{fior}), so that the needed
systematic error would be even larger than what we have considered for
the production channel.

Thus nuclear solutions do not seem allowed by the current data.
Of course, it remains possible that extremely large
(factors $\ga 100$) systematic errors lurk in otherwise
negligible \li7 production and destruction channels
\cite{cva}.  For these reasons, continued
efforts to improve nuclear cross section experiments and theory (with
particular attention to absolute normalizations and systematics) will
reap benefits for BBN as well as solar neutrinos.  Tighter
experimental errors (including systematics) will reduce the BBN
theoretical uncertainty budget, which will not only further clarify
the seriousness of the \li7 problem, but also allow for stronger
constraints on astrophysics \cite{cfo3} when and if the \li7 problem
is resolved.  In this respect, we particularly call attention to the
$\he3(\alpha,\gamma)\be7$ reaction, but also to $\be7(p,\gamma)\bor8$,
as they are undoubtedly linked through solar neutrinos.  Determining a
more accurate low-energy extrapolation in either of these reactions
will impact the other through the solar neutrino constraint on the
parameter $x=S_{17}S_{34}^{0.81}$.

Where, then, does the \li7 problem stand?  We have found nuclear
reaction systematics are very unlikely to be the dominant source of
the discrepancy. Of the remaining possibilities, the most conservative
is that the problem is dominated by systematic errors in the
observational \li7 value.  This could either be due to difficulties in
the understanding the stellar parameters and in extracting the
abundance from spectral lines, or from stellar evolution effects which
deplete Li without introducing large dispersion in the Spite plateau.
A similarly conventional solution would ascribe the \li7 discrepancy
to a combination of nuclear and observational systematics, both at the
edge of what is currently allowed.

Finally, a more radical but intriguing possibility would be that new
physics is required.  If this is so, nature has been somewhat subtle
in revealing this twist, as the perturbation to standard BBN has
been small enough not to be noticed until now.\footnote{ If so, this
probably has been fortuitous for the development of cosmology.  Had
there always been large problems with standard BBN, one can imagine
that this would have led to great skepticism about the viability of
the hot big bang framework.} Nonstandard scenarios have
already been proposed to alleviate the \li7 problem by introducing new
physics, e.g., by a late-decaying gravitino \cite{cefo}.  
However, most of the scenarios require
fine tuning, as one wishes to reduce \li7 without spoiling the superb
concordance between deuterium and the CMB.

In summary, we use solar neutrinos to remove the possibility of a
solution to the \li7 problem from the $\he3(\alpha,\gamma)\be7$
reaction, and thereby cast more doubt that the problem is due to
nuclear systematics.  By removing a possible resolution, we have both
clarified the problem, and made it more acute.  In our view, the most
important arena now is the observations and astrophysics which lead to
the primordial \li7 inference.  And while we continue to suspect that
this is the likely solution, a parallel examination of nonstandard BBN
scenarios is at this point not unwise.

\section*{Acknowledgments}

We thank John Bahcall,  Barry Davids, Byron Jennings, and Vijay Pandharipande for
useful discussions.  The work of K.A.O. was partially supported by DOE
grant DE-FG02-94ER-40823. The work of B.D.F. and R.H.C. was supported
by the National Science Foundation under grant AST-0092939.


\begin{thebibliography}{0}

\bibitem{wmap}
C.L.~Bennett {\it et al.}, {\it Astrophys.~J.} submitted,
arXiv:astro-ph/0302207;
D.~N.~Spergel {\it et al.},
{\it Astrophys.~J.} submitted,
arXiv:astro-ph/0302209;

\bibitem{bbn} T. P. Walker, G. Steigman, D. N. Schramm, 
 K. A. Olive and K. Kang, {\it Ap.J.} {\bf 376} (1991) 51;
K. A. Olive, G. Steigman, and T. P. Walker, {\it Phys. Rep.} {\bf 333} (2000) 389; 
B. D. Fields and S. Sarkar, 
{\it Phys.\ Rev.} {\bf D66} (2002) 010001.

\bibitem{sarkar}
S. Sarkar, {\it Rep. Prog. Phys.} {\bf 59} (1996) 1493.

\bibitem{fo} B.D. Fields and K.A. Olive, {\it Phys. Lett.} {\bf B368}
(1996) 103; \\ B.D. Fields, K. Kainulainen, D. Thomas, and K.A. Olive,
{\it New Astronomy} {\bf 1} (1996) 77.


\bibitem{lik}
N.~Hata, R.~J.~Scherrer, G.~Steigman, D.~Thomas, T.~P.~Walker, S.~Bludman and
P.~Langacker,
Phys.\ Rev.\ Lett.\  {\bf 75}, 3977 (1995)
[arXiv:hep-ph/9505319];
N.~Hata, G.~Steigman, S.~Bludman and P.~Langacker,
Phys.\ Rev.\ D {\bf 55}, 540 (1997)
[arXiv:astro-ph/9603087];
S.~Esposito, G.~Mangano, G.~Miele and O.~Pisanti,
Nucl.\ Phys.\ B {\bf 568} (2000) 421
[arXiv:astro-ph/9906232];
S.~Burles, K.~M.~Nollett and M.~S.~Turner,
Astrophys.\ J.\  {\bf 552}, L1 (2001)
[arXiv:astro-ph/0010171].

\bibitem{fior}
G.~Fiorentini, E.~Lisi, S.~Sarkar and F.~L.~Villante,
simulations,''
Phys.\ Rev.\ D {\bf 58}, 063506 (1998)
[arXiv:astro-ph/9803177];


\bibitem{cfo1}
R.~H.~Cyburt, B.~D.~Fields and K.~A.~Olive,
New Astron.\  {\bf 6} (1996) 215
[arXiv:astro-ph/0102179].

\bibitem{cbi}
J.~L.~Sievers {\it et al.},
Comparisons with BOOMERANG, DASI, and MAXIMA,''
arXiv:astro-ph/0205387.

\bibitem{acb}
J.~H.~Goldstein {\it et al.},
arXiv:astro-ph/0212517.

\bibitem{2df}
W.~J.~Percival  {\it et al.},
{\it Monthly Not.~Royal Astr.~Soc.},
{\bf 337}, 1297 (2001).


\bibitem{cfo3}
R.~H.~Cyburt, B.~D.~Fields and K.~A.~Olive,
Phys.\ Lett.\  {\bf 567} (2003) 227
[arXiv:astro-ph/0302431].

\bibitem{bt} S. Burles and D. Tytler, {\it Ap.J.} {\bf 499}, 699 (1998);
{\it Ap.J.} {\bf 507}, 732 (1998).

\bibitem{omeara}
J.~M.~O'Meara, D.~Tytler, D.~Kirkman, N.~Suzuki, J.~X.~Prochaska, D.~Lubin and
A.~M.~Wolfe,
Astrophys.\ J.\  {\bf 552}, 718 (2001)
[arXiv:astro-ph/0011179].

\bibitem{kirkman}
D.~Kirkman, D.~Tytler, N.~Suzuki, J.~M.~O'Meara and D.~Lubin,
arXiv:astro-ph/0302006.

\bibitem{pet}
M.~Pettini and D.~V.~Bowen,
Convergence with the Baryon Density from the CMB?,''
Astrophys.\ J.\  {\bf 560}, 41 (2001)
[arXiv:astro-ph/0104474].


\bibitem{iz}
Y.~I.~Izotov, T.~X.~Thuan, and V.~A.~Lipovetsky,
Ap.\ J. {\bf 435} (1994) 647;
Y.~I.~Izotov, T.~X.~Thuan, and V.~A.~Lipovetsky,
Ap.\ J. {\bf 108} (1997) 1;
Y.~I.~Izotov and T.~X.~Thuan,
Ap.\ J. {\bf 500} (1998) 188;
Y.~I.~Izotov and T.~X.~Thuan,
astro-ph/0310421.

\bibitem{oss}
K.A. Olive, E. Skillman, and G. Steigman, Ap.J.
{\bf 483} (1997) 788.

\bibitem{fo98}
B.~D.~Fields and K.~A.~Olive,
{\it Ap.\ J.} {\bf 506} (1998) 177
[arXiv:astro-ph/9803297].


\bibitem{peim} 
M. Peimbert, A. Peimbert and M. T. Ruiz,
Ap.\ J. {\bf 541} (2000) 688;
A. Peimbert, M. Peimbert and V. Luridiana, Ap.\ J. {\bf 565} (2002) 668.

\bibitem{OSk} K.A. Olive, and E. Skillman, {\it New Ast.}
{\bf 6} (2001) 119
[arXiv:astro-ph/0007081];
D. Sauer, and K. Jedamzik, {\it A.A.} {\bf 381}, 361 (2002);
R.~Gruenwald, G.~Steigman and S.~M.~Viegas,
{\em Astrophys.~J.} {\bf 567} (2002) 931.
[arXiv:astro-ph/0109071].

\bibitem{spites}
F. Spite, and M. Spite, A.A.  {\bf 115} (1982) 357.

\bibitem{Nollett:2001ub}
K.~M.~Nollett,
Phys.\ Rev.\ C {\bf 63} (2001) 054002
[arXiv:nucl-th/0102022].


\bibitem{bah}
Bahcall, J.N., Pinsonneault, M.H., \& Basu,\ S. ApJ {\bf 555} (2001) 990.

\bibitem{adelberger}
Adelberger, E.C. {\em et al}.\ Rev. Mod. Phys. {\bf 70} (1998) 1265.

\bibitem{nacre}
C.~Angulo et al.~(NACRE Collaboration), 
{\it Nuc.\ Phys.\ A} {\bf 656} (1999) 3.


\bibitem{nb} 
K.~M.~Nollett and S.~Burles,
{\it Phys.\ Rev.}  {\bf D61} (2000) 123505
[arXiv:astro-ph/0001440].

\bibitem{cva}
A.~Coc, E.~Vangioni-Flam, P.~Descouvemont, A.~Adahchour and C.~Angulo,
arXiv:astro-ph/0309480.


\bibitem{rbofn} S.G. Ryan,  T.C. Beers, K.A. Olive, B.D. Fields, and J.E.
Norris, {\it Ap.J. Lett.} {\bf 530}, L57 (2000).

\bibitem{rnb} S.G. Ryan,  J.E. Norris, and T.C. Beers,
Ap.\ J.\  {\bf 523} (1999) 654.

\bibitem{sneden}
C. Sneden, et al.,
Ap.\ J.\  {\bf 591} (2003) 936.

\bibitem{ddk}
C.P. Deliyannis,  P. Demarque, and S.D. Kawaler, 
Ap.\ J.\ Supp.\ ,{\bf 73} (1990)  21.

\bibitem{dep}
S. Vauclair,and C. Charbonnel, Ap. J. {\bf 502} (1998) 372;
M.~H.~Pinsonneault, T.~P.~Walker, G.~Steigman and V.~K.~Narayanan,
Ap. J. {\bf 527} (1998) 180
[arXiv:astro-ph/9803073];
M.~H.~Pinsonneault, G.~Steigman, T.~P.~Walker, and V.~K.~Narayanan,
Ap. J. {\bf 574} (2002) 398
arXiv:astro-ph/0105439.

\bibitem{bon1}
P.~Bonifacio, et al., 
Astron.~Astrophys., {\bf 390} (2002) 91.

\bibitem{bon2}
P. Bonifacio,
Astron.\ Astrophys.\ {\bf 395} (2002) 515.

\bibitem{bbnwoc}
Esmailzadeh, R., Starkman, G.D., \& Dimopoulos, S. ApJ {\bf 378} (1991) 504.

\bibitem{sno}
SNO Collaboration,
submitted to Phys. Rev. Lett.\ (2003)
arXiv:nucl-ex/0309004


\bibitem{bahcall} 
J.~N. Bahcall, Neutrino Astrophysics (1989),
Cambridge and New York, Cambridge University Press.

\bibitem{bahulr}
Bahcall, J.~N.~\& Ulrich, R.~K.\  Rev. Mod. Phys., {\bf 60} (1988) 297.

\bibitem{junghans}
Junghans, A.R. {\em et al}.
arXiv:nucl-ex/0308003

\bibitem{davids}
Davids, B. \& Typel, S.,
Phys. Rev. C {\bf 68} (2003) 045802
arXiv:nucl-ex/0304054

\bibitem{cefo}
S.~Dimopoulos, R.~Esmailzadeh, L.~J.~Hall and G.~D.~Starkman,
Astrophys.\ J.\  {\bf 330} (1988) 545;
R.~H.~Cyburt, J.~R.~Ellis, B.~D.~Fields and K.~A.~Olive,
Phys.\ Rev.\ D {\bf 67}, 103521 (2003)
[arXiv:astro-ph/0211258].

\end{thebibliography}
\end{document}